\documentclass[twocolumn]{aa}
\usepackage{psfig,epsf,graphicx,lscape}
\begin{document}

  \title{Optical and X-ray Rest-frame Light Curves of the BAT6 sample}

   \author{A. Melandri\inst{1}, S. Covino\inst{1}, D. Rogantini\inst{1,2}, R. Salvaterra\inst{3}, B. Sbarufatti\inst{1}, M. G. Bernardini\inst{1}, S. Campana\inst{1}, P. D'Avanzo\inst{1}, V. D'Elia\inst{4,5}, D. Fugazza\inst{1}, G. Ghirlanda\inst{1}, G. Ghisellini\inst{1}, L. Nava\inst{5}, S. D. Vergani\inst{6,1}, and G. Tagliaferri\inst{1}
          }


   \institute{$^{1}$INAF - Osservatorio Astronomico di Brera, Via Bianchi 46, I-23807 Merate (LC), Italy\\
                   $^{2}$Universit\`{a} degli Studi di Milano-Bicocca, Piazza della Scienza 3, I-20126, Milano, Italy\\
                   $^{3}$INAF - IASF Milano, Via E. Bassini 15, I-20133, Milano, Italy\\
                   $^{4}$ ASI - Science Data Center, Via del Politecnico snc, I-00133 Roma, Italy\\
   		$^{5}$ INAF - Osservatorio Astronomico di Roma, Via Frascati 33, I-00040, Monte Porzio Catone (RM), Italy\\
                   $^{6}$APC, Univ. Paris Diderot, CNRS/IN2P3, CEA/Irfu, Obs. de Paris, Sorbonne Paris Cit\'{e}, France\\
                   $^{7}$GEPI, Observatoire de Paris, CNRS, Univ. Paris Diderot, 5 place Jule Janssen, F-92190, Meudon, France\\
              email: andrea.melandri@brera.inaf.it
             }

   \date{}

\abstract
{}
{We present the rest-frame light curves in the optical and X-ray bands of an unbiased and complete sample of Swift long Gamma-Ray Bursts (GRBs), namely the BAT6 sample.}
{The unbiased BAT6 sample (consisting of 58 events) has the highest level of completeness in redshift ($\sim 95 \%$), allowing us to compute the rest-frame X-ray and optical light curves for 55 and 47 objects, respectively. We compute the X-ray and optical luminosities accounting for any possible source of absorption (Galactic and intrinsic) that could affect the observed fluxes in these two bands.}
{We compare the behaviour observed in the X-ray and in the optical bands to assess the relative contribution of the emission during the prompt and afterglow phases. We unarguably demonstrate that the GRBs rest-frame optical luminosity distribution is not bimodal, being rather clustered around the mean value Log(L$_{\rm R}$) = 29.9 $\pm$ 0.8 when estimated at a rest frame time of 12 hr. This is in contrast with what found in previous works and confirms that the GRB population has an intrinsic unimodal luminosity distribution. For more than 70$\%$ of the events the rest-frame light curves in the X-ray and optical bands have a different evolution, indicating distinct emitting regions and/or mechanisms. The X-ray light curves normalised to the GRB isotropic energy ($E_{\rm iso}$), provide evidence for X-ray emission still powered by the prompt emission until late times ($\sim$ hours after the burst event). On the other hand, the same test performed for the $E_{\rm iso}$-normalised optical light curves shows that the optical emission is a better proxy of the afterglow emission from early to late times.}
{}

   
\authorrunning{Melandri et al. 2014}
\titlerunning{Rest-frame light curves of the complete BAT6 sample}

   \maketitle


%

\section{Introduction}

The advance in capability to detect, fast re-point and observe for long time intervals, Gamma-Ray Bursts (GRBs) by the {\it Swift} satellite (\cite{geh}) has allowed in the last few years the creation of uniform collections of events for statistical purposes (\cite{nousek,zhang,geh2,evans0,roming,oates,kann,rykoff,nyse,li,mela0,kann2,liang2,ruben,raffa}). In particular, depending on the selecting criteria defining the sample, these collections of GRBs would probe in different manner the parameters space of rest-frame time-brightness. Among all the others, the more complete sample of bright {\it Swift} GRBs is the one by Salvaterra et al. (2012), where the selection of bright events in the $\gamma$-ray band led to an homogeneous set of GRBs, with very good coverage of the X-ray and optical bands, giving also a completeness in redshift  of $\sim 95\%$.

The exploitation of this complete sample gave for the first time some interesting statistical results of the overall class of bright GRBs: a) a strong evolution of the luminosity or density distribution is needed to account for the observations (\cite{ruben}); b) selection effects are negligible in shaping the spectral-energy correlations that might be due to a strong physical mechanism common to large majority of GRBs (\cite{nava}; \cite{ghirla}); c) the existence of a true population of dark GRBs  that generate in much denser environments (\cite{mela}); d) a strong correlation between GRB darkness and X-ray absorbing column densities (\cite{campana}); e) a significant correlation between X-ray luminosity and prompt $\gamma$-ray energy and luminosity (\cite{davanzo}); f) a distribution of rest-frame absorption coefficients peaked at low values with a tail of dark GRB highly absorbed (\cite{covino}).

In this paper we finally investigate the properties of the rest-frame optical luminosity of the complete BAT6 sample (details about the selecting criteria of the sample in \cite{ruben}), showing and comparing the rest-frame light curves of each event in the  $\gamma$-rays, X-ray and optical bands. Throughout the paper we assume a standard cosmology with $H_{\rm 0}$ = 72 km s$^{-1}$ Mpc$^{-1}$, $\Omega_{\rm m}$ = 0.27, and $\Omega_{\rm \Lambda}$ = 0.73.



\section{Luminosity light curves fitting}

We followed the classification scheme described in Margutti et al. (2013), that divided the light curves in four main classes: the class {\bf 0} for events decaying with simple power-law, the class {\bf Ia} ({\bf Ib}) for a smooth broken power-laws that go from a shallower (steeper) to a steeper (shallower) decay, the class {\bf IIa} or {\bf IIb} for events that display steep-shallow-steep (canonical) or shallow-steep-shallow decays respectively, and finally the class {\bf III} for those events that show a further late time break in their light curves. Margutti et al. (2013) did not take into account of any possible flare or density bump in their fit, concentrating their classification to the underlying smooth components. We instead tried to model independently the luminosity X-ray light curve (rest-frame) without excluding any data point from our fit and flagging the events with flares and/or bumps with a suffix {\bf F} or {\bf B}\footnote{We considered as flares powerful and fast re-brightenings superposed to the normal decay at early times, while bumps are much broad features happening at later times.}. In the optical band we converted the observed data into rest-frame luminosity, correcting for any intrinsic absorption, with the aim of comparing it with the X-ray behaviour. The classification scheme remains the same as for the X-ray band, with the addition of the suffix {\bf Onset} for those events that display a clear peak rising with $\alpha \leq -2.0$ (where the afterglow emission is $\propto t^{-\alpha}$) and having a peak time $t_{\rm peak} < 3 \times 10^{2}$~s.

We performed the fit of each luminosity light curve singularly, trying to model any possible additional component (flare or bump) with respect to the underlying prompt-afterglow emission in the X-ray and optical bands. For two cases, namely GRB\,060614 and GRB\,091127, we exclude from our fit the late time optical detections related to the contribution of the underlying host galaxy and/or supernova. Moreover, for the first of these two events, we did not fit also the early time X-ray data, not reproducible with a simple power-law decay (\cite{mangano}). Our classification, reported in Table \ref{tabtypes}, might therefore differ for some events from the one found by Margutti et al. (2013) and Evans et al. (2009), that automatically executed their fitting procedures over a much larger sample of GRBs, sometimes using only power-law segments to model the light curves decays.


\begin{figure*}
\includegraphics[width=18cm,height=21cm]{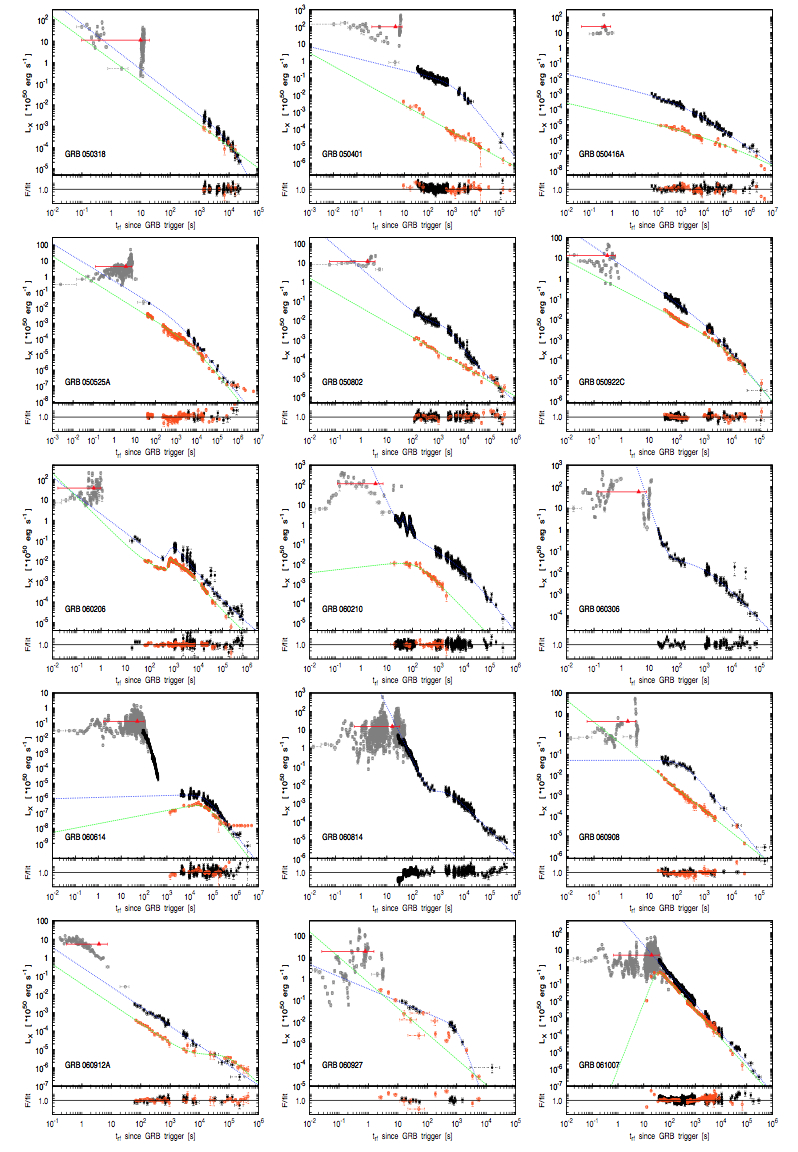} 
 \caption{Rest frame $\gamma$-ray (grey), X-ray (black) and optical (orange) luminosity. The red triangle shows the value of average isotropic $\gamma$-ray luminosity L$_{\rm iso}$. Blue and green dotted lines represent the light curve best fit for the X-ray and optical bands, respectively.}
\label{FigLCs1}
\end{figure*}

\addtocounter{figure}{-1}

\begin{figure*}
\includegraphics[width=18cm,height=21cm]{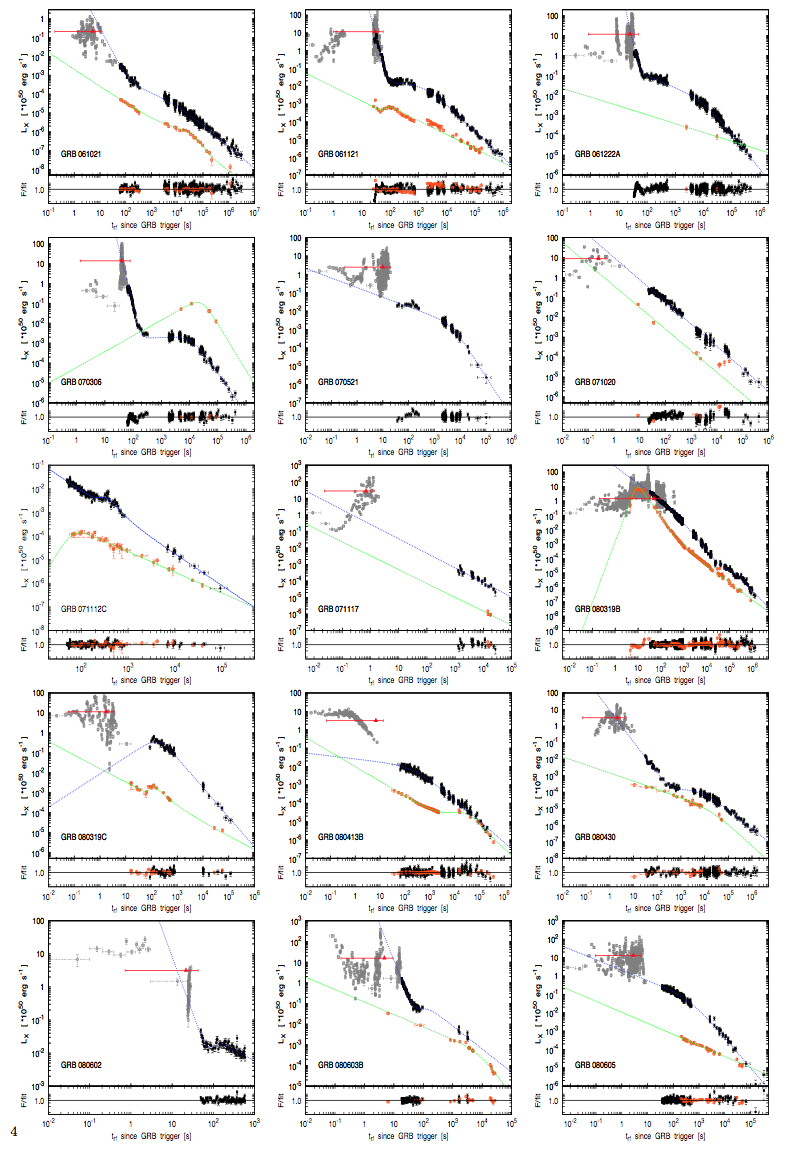} 
 \caption{-continued.}
\vspace{0cm}
\label{FigLCs2}
\end{figure*}

\addtocounter{figure}{-1}

\begin{figure*}
\includegraphics[width=18cm,height=21cm]{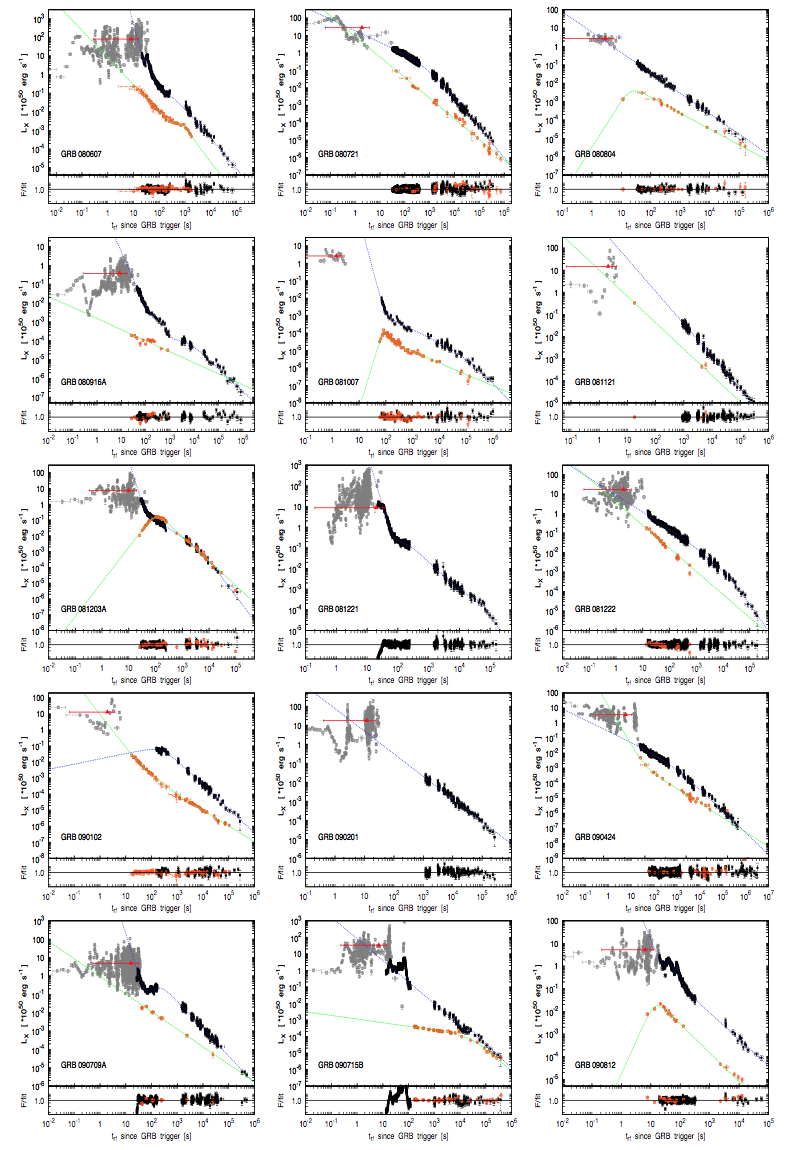} 
 \caption{-continued.}
\vspace{0cm}
\label{FigLCs2}
\end{figure*}

\addtocounter{figure}{-1}

\begin{figure*}
\includegraphics[width=18cm,height=18cm]{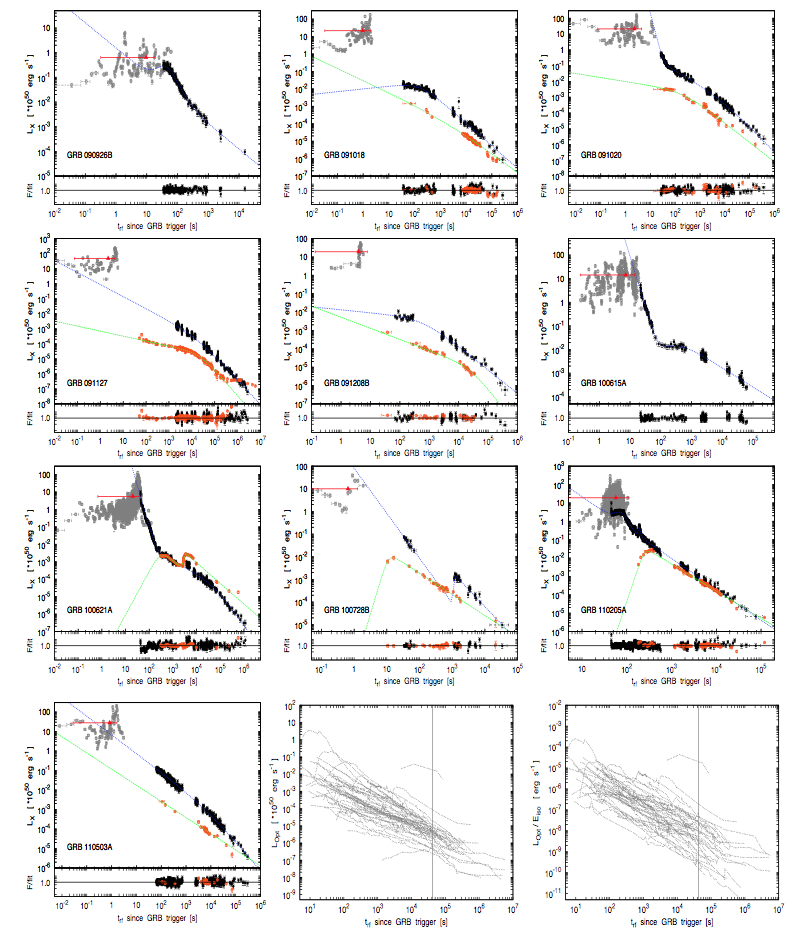} 
 \caption{-continued.}
\vspace{0cm}
\label{FigLCs2}
\end{figure*}


\section{Results}

In Fig. \ref{FigLCs1}  we show the $\gamma$-ray (grey), X-ray (black) and optical (orange) rest-frame luminosity for all the GRBs with redshift of the complete sample. For each event we retrieved data in the $\gamma$-ray and in the X-ray bands from the online {\it Swift} Burst Analyser (\cite{evans}); {\it Swift}-BAT data have been rescaled to the {\it Swift}-XRT [0.3-10]~keV energy range. The optical data were collected from all the available sources (published dedicated papers and/or GCNs). We then converted all the observed $\gamma$-ray and X-ray data into rest-frame luminosity in the [2-10]~keV energy range (L$_{\rm BAT}$ and L$_{\rm XRT}$) and optical data into rest-frame luminosity in the $U$-band (L$_{\rm Opt}$). For the optical conversion we took into account the spectral and absorption parameters of each single burst from the detailed analysis of Covino et al. (2013) (see next section). The results of the light curves fitting in the X-ray and optical bands are reported in Table \ref{tablc} and  \ref{tablc2}, respectively.

In the X-ray band 20$\%$ of the luminosity light curves are represented by a single power-law decay (class 0), 42$\%$ by a single break light curve ($\sim$36$\%$ from shallow to steep, class Ia, and $\sim$6$\%$ from steep to shallow decays, class Ib) and the remaining 38$\%$ by a double break light curve ($\sim$36$\%$ by the canonical steep-shallow-steep decay, class IIa, and $\sim$2$\%$ by the shallow-steep-shallow decay, class IIb). In the optical band the behaviour is different, with the majority of the events (49$\%$) belonging to the class 0, 31$\%$ described by single break light curve ($\sim$24$\%$ of class Ia and $\sim$7$\%$ of class Ib), 6$\%$ by the canonical double break light curve (class IIa) and for the remaining 14$\%$ (8 out of 55) no optical data are available to assess their light curve behaviour.

 \begin{table}
       \caption[]{Number of GRBs per light curve type. Numbers in parentheses are the events for that class that display the onset of the optical afterglow. We note that the class III defined by Margutti et al. (2013) is never seen in the complete sample.}
         \label{tabtypes}
 \centering
  \begin{tabular}{ccccccc}
   \hline
    & Total & 0 & Ia & Ib & IIa & IIb \\
  \hline
   X-rays & 55 & 11 & 20 & 3 & 20 & 1 \\
   Optical & 47 & 27 (8) & 13 & 4 (2) & 3 & 0 \\
  \hline
  \end{tabular}
\end{table}

\subsection{Rest-frame optical luminosity}

The majority of the optical observations for the GRBs in the sample were carried out in the optical $R$ and $I$ bands. Since the median redshift of our complete GRBs sample is {\it z} $\sim$ 1.6 (\cite{ruben}) these observational bands would correspond to a rest-frame frequency $\nu_{\rm RF} \sim 0.9 \times 10^{15}$ Hz. This value is close to the $U$-band rest frame frequency ($\nu_{\rm RF}\approx \nu_{\rm U} = 0.82 \times 10^{15}$ Hz). We therefore reported, after taking into account the spectral ($\beta_{\rm O}$) and rest-frame absorption (A$_{V}$) parameters (\cite{covino}), the observed optical/IR flux of each burst to the correspondent rest-frame flux at the frequency $\nu_{\rm U}$ and then converted it into rest-frame luminosity\footnote{For GRB\,071117 we assumed $\beta_{\rm O}$ = 1.0 since, due to the paucity of optical data, it was not possible to better constrain the spectral index.} (L$_{\rm opt}$, shown in Fig. \ref{FigLCs1}). 

In general, if a GRB is mainly observed in the optical band $\nu_{\rm obs}$, then the Galactic corrected observed flux ($f_{\rm obs}$) will be reported to the rest-frame optical luminosity (L$_{\rm opt}$) at the frequency $\nu_{\rm U}$ applying the formula

\begin{eqnarray}
L_{\rm opt} = 4 \pi d_{\rm L}^{2} \times 10^{0.4A_{\rm U}} \times f_{\rm obs} \times \left[\frac{\nu_{\rm U}}{(1+{\it z}) \nu_{\rm obs}}\right]^{-\beta_{\rm O}}
\end{eqnarray}

\noindent where $10^{0.4A_{\rm U}}$ is the absorption correction, $d_{\rm L}$ is the luminosity distance and $z$ is the redshift of the GRB. 

\begin{figure}
\includegraphics[width=9cm,height=8cm]{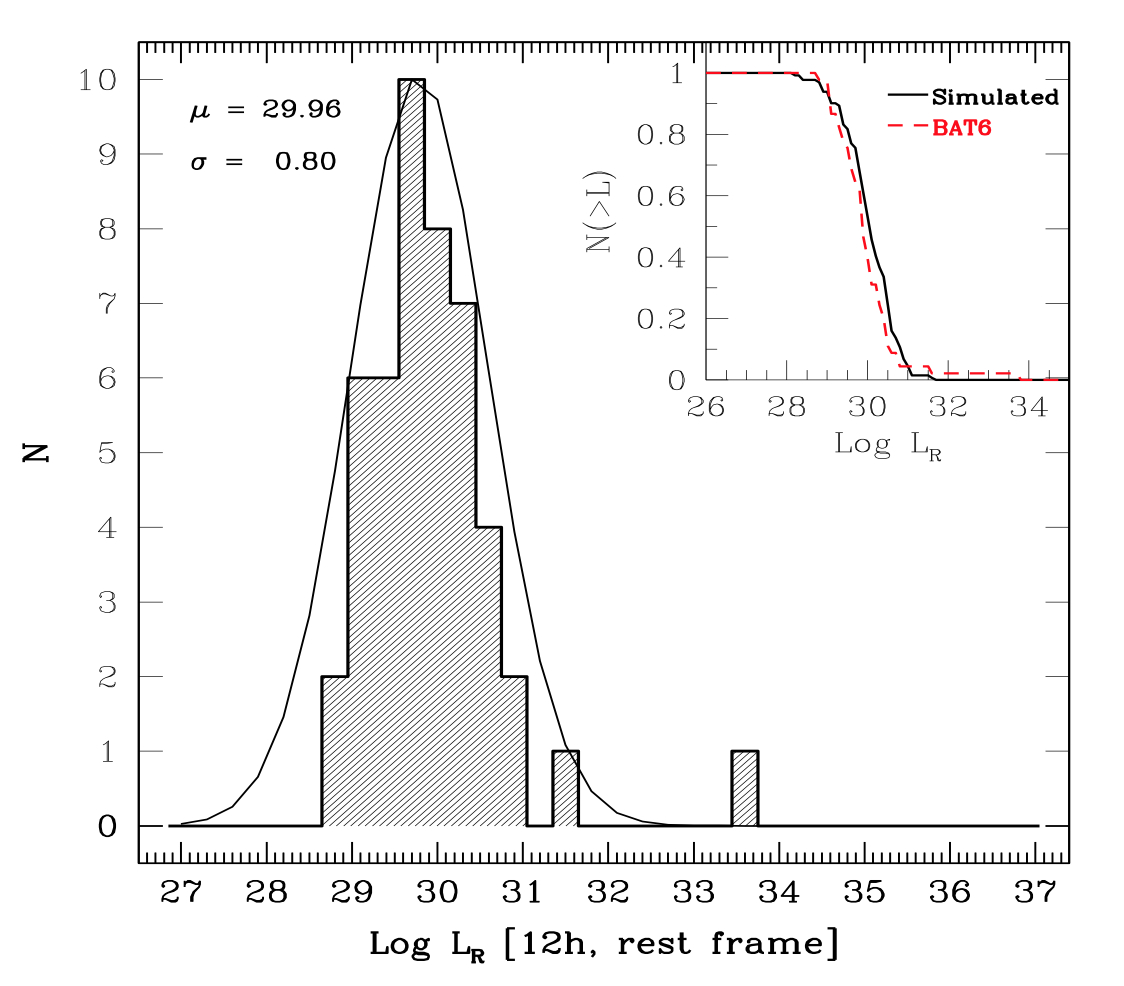}
\caption{Optical rest-frame luminosity ($R$-band, erg s$^{-1}$ Hz$^{-1}$) distribution at $t_{\rm RF}$ = 12 hr. The mean value of the distribution is $\mu =$ 29.9, with a dispersion $\sigma=$ 0.8. {\it Inset}: cumulative distribution of the optical rest-frame luminosity for the BAT6 sample (red) and for a simulated population of GRBs obtained with the PSYCHE code (\cite{ghirla2}). }
\label{FigoptL}
\end{figure}

The result for each single burst is reported in one single panel of Fig. \ref{FigLCs1}, where the optical behaviour is compared with the rest-frame luminosity in $\gamma$ and X-rays. In Fig. \ref{FigoptL} we show the $R$-band rest-frame optical luminosity, as estimated from equation 1, for the GRBs in our sample, evaluated at $t_{\rm rf}= 12$~hr in the rest-frame. Considering the median redshift of the BAT6 sample, with that choice we are considering a time in the observer frame larger than 1~day after the burst event. At such late times, the observed optical luminosity should come only from one component, i.e. the afterglow. As it can be seen in Fig. \ref{FigoptL} the bimodality of the luminosity distribution at that time claimed in previous works (e.g. \cite{liang}, \cite{nardini1}, \cite{kann}, \cite{nardini2}) is not confirmed with our complete sample. There are no clear gaps in the luminosity range $28 \le$ Log (L$_{\rm R}$) $\le 30$. Only one event has an extremely bright optical luminosity; this event correspond to GRB\,070306, the event with the highest rest-frame optical absorption also found by Covino et al. (2013). The mean value of the L$_{\rm R}$ distribution is $\mu =$ 29.96 ($\sigma =$ 0.80) erg s$^{-1}$ Hz$^{-1}$. Assuming a central value for the $R$-band of $\nu_{\rm R} = 4.6 \times 10^{14}$~Hz, the mean value of the L$_{\rm R}$ distribution of the BAT6 sample corresponds to $\mu =$ 44.6 erg s$^{-1}$, consistent with the value found in other works ($\mu =$ 44.20, $\sigma =$ 0.67 by \cite{mela0}; $\mu =$ 44.50, $\sigma =$ 0.74 by \cite{elena}).

\begin{figure}
\includegraphics[width=9cm,height=11.0cm]{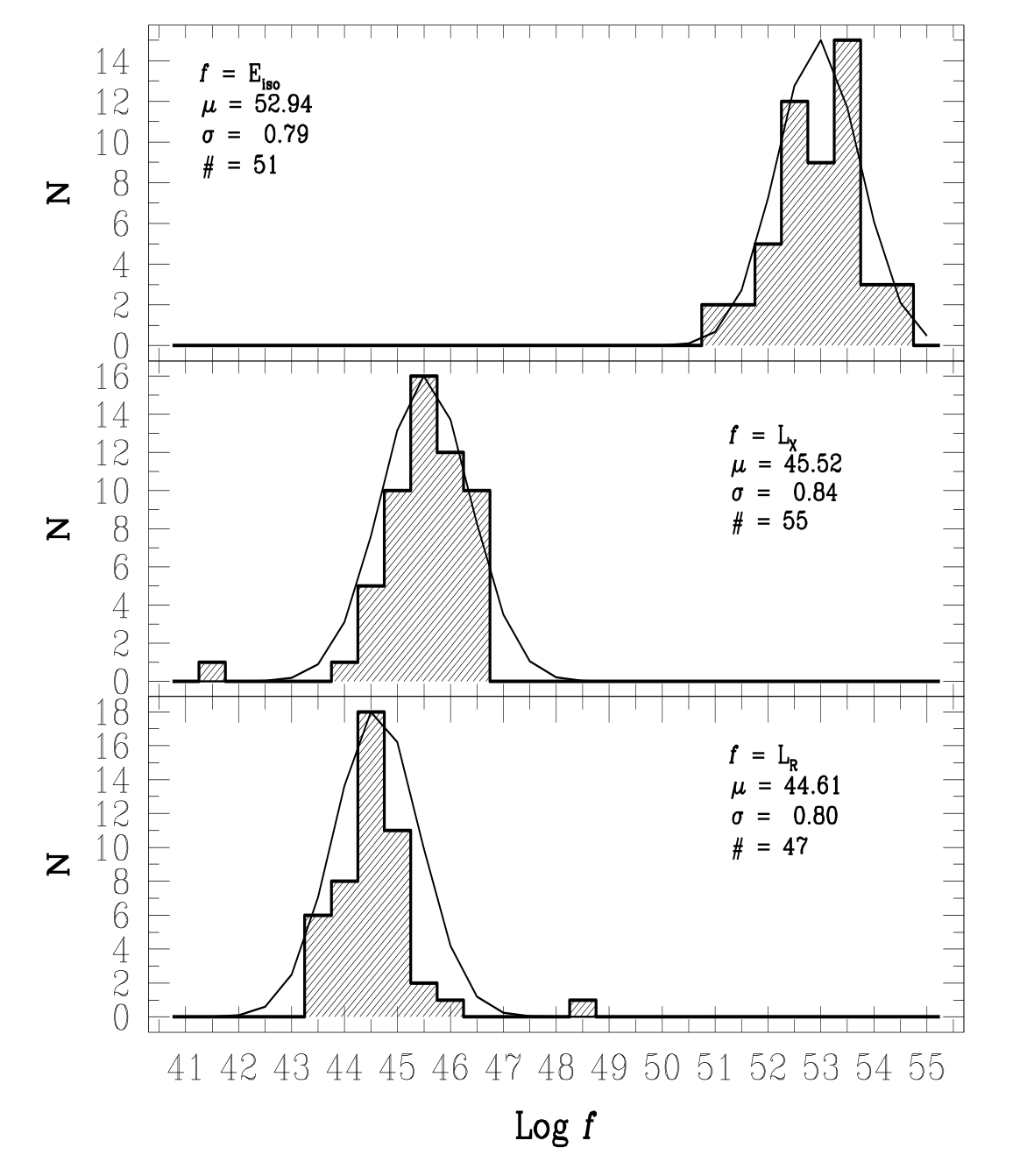}
 \caption{Distributions of E$_{\rm iso}$ (top), L$_{\rm XRT}$ (mid) and L$_{\rm opt}$ (bottom) of the BAT6 sample for all the events with a confirmed redshift. For each histogram we report the number of events considered ($\#$), the mean value ($\mu$) and dispersion ($\sigma$) of the distribution. Units on the x-axis {from top to bottom} are $erg$, $erg~s^{-1}$ and $erg~s^{-1}$, respectively.}
\label{FigoptL1}
\end{figure}
\begin{figure}
\includegraphics[width=9cm,height=10.0cm]{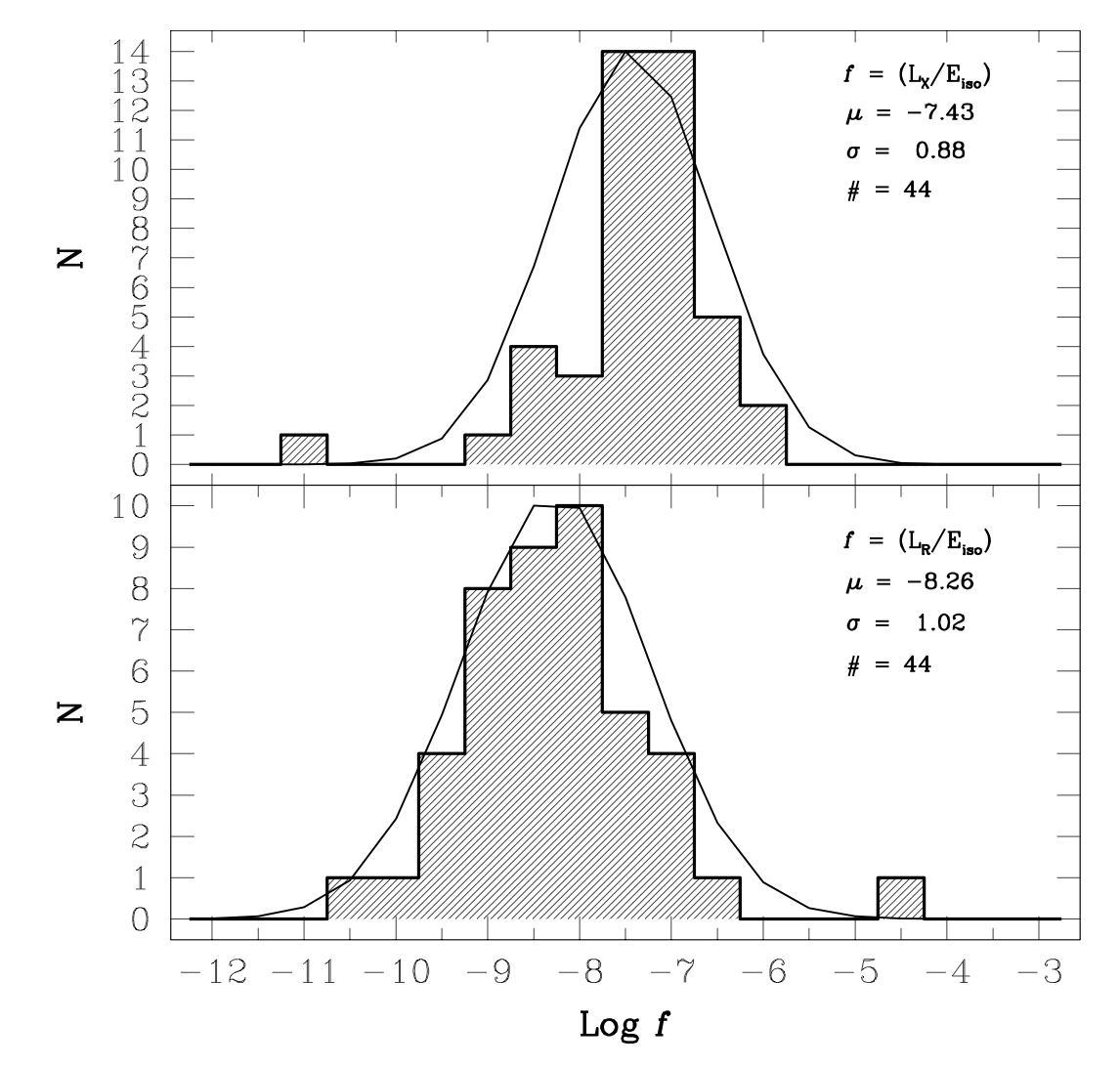}
 \caption{E$_{\rm iso}$-normalised L$_{\rm XRT}$ (top) and L$_{\rm opt}$ (bottom). The latter has a larger dispersion than the former.}
\label{FigoptL2}
\end{figure}

The overall distribution of the L$_{\rm opt}$ for our complete sample (Fig. \ref{FigLCs1}, second to last panel) spans $\sim$4 order of magnitudes from very early time up to days after the burst event. This trend is mimicked also by the distribution of the E$_{\rm iso}$-normalised optical luminosity, i.e. the luminosities normalised to their isotropic energies (Fig. \ref{FigLCs1}, last panel), where the dispersion at late times is even larger. In particular, no strong clustering at early time is found for L$_{\rm opt}$/E$_{\rm iso}$ and the scatter of the distribution remains nearly constant for the entire duration of the optical observation. This is in contrast to what found by D'Avanzo et al. (2012) where the distribution of L$_{\rm X}$ was strongly correlated with the isotropic energy. In Fig. \ref{FigoptL1} a direct comparison between the distributions of E$_{\rm iso}$, L$_{\rm XRT}$ and L$_{\rm opt}$ of the BAT6 sample is shown: in that figure we considered all the events with confirmed redshift for which the estimate of one of the three variable has been possible. In Fig. \ref{FigoptL2} we show how the E$_{\rm iso}$-normalised optical luminosity (L$_{\rm R}$/E$_{\rm iso}$; bottom panel) has a larger distribution with respect to the E$_{\rm iso}$-normalised X-ray luminosity (L$_{\rm X}$/E$_{\rm iso}$; top panel) already described in D'Avanzo et al. (2012). This result strengthens the idea that the X-ray luminosity is a better proxy of the prompt emission while the optical luminosity describes more accurately the afterglow emission.

The agreement between the $\gamma$-ray (L$_{\rm BAT}$) and X-ray (L$_{\rm XRT}$) light curves is quite good, where typically the latter seem to be a natural continuation of the former when there is still some overlapping in time (i.e. for $t_{\rm rf}<10^{2}$~s). This, coupled with the the fact that the light curves seem to have different behaviours in the X-ray and in the optical bands for the majority of the GRBs, might indicate that, at least at very early times, the emission in the X-ray is still contaminated by the prompt emission of the GRB, while at later times afterglow emission dominates in all bands (i.e. \cite{ghise}). For some cases the back extrapolation of L$_{\rm XRT}$ seems to under-estimate (i.e. GRB\,050416A and GRB\,071117) or over-estimate (i.e. GRB\,061121) the L$_{\rm BAT}$. This might be only a visual effect due to the fact that for these events there is a gap between the last emission detected in $\gamma$-ray (sometimes the emission in $\gamma$-ray does not last more then few tenth of seconds) and the first detection in X-ray (sometimes observation in X-ray will not start before few hundreds of seconds).

\subsection{Cumulative distribution}

In the inset of Fig. \ref{FigoptL} we compare the cumulative distribution of the observed optical luminosity for the BAT6 sample with the simulated cumulative distribution obtained with the PSYCHE code (\cite{ghirla2}). This code generates a synthetic population of GRBs that reproduces some properties of the real population of GRBs observed by {\it Swift}, {\it Fermi} and BATSE (e.g., flux and fluence distributions, rate of detected events and rest-frame $E_{\rm p}$-$E_{\rm iso}$ correlation). Then, it calculates the flux (luminosity) at any given time and frequency. Therefore, starting from the observed properties of the BAT6 sample, we estimated the expected optical luminosity at $t_{\rm RF}$ = 12 hr. As it can be seen in Fig. \ref{FigoptL} (inset), the simulated cumulative distribution (black solid line) agrees very well with the luminosity distribution observed for the BAT6 sample (red dashed line). The PSYCHE code is here used to compare the distribution of the simulated burst optical fluxes with the real ones. The comparison of the X-ray flux distribution is beyond the scope of the present paper and will be included in a forthcoming dedicated paper.

The properties of the BAT6 sample can be reproduced fixing the index of the expected power-law electron energy distribution $p$ = 2.5, the ratio between the energy of non-thermal electrons and the energy dissipated at the shock $\epsilon_{\rm e}$ = 0.02, the ratio between the energy gained by the magnetic field and the energy dissipated at the shock $\epsilon_{\rm B}$ = 0.008, and assuming that the parameter describing the density of the medium ($n$) has a uniform distribution between 0.1 and 30 cm$^{-3}$ (\cite{ghirla2}).

\subsection{Statistics}

\begin{figure*}
\includegraphics[width=18cm,height=15.0cm]{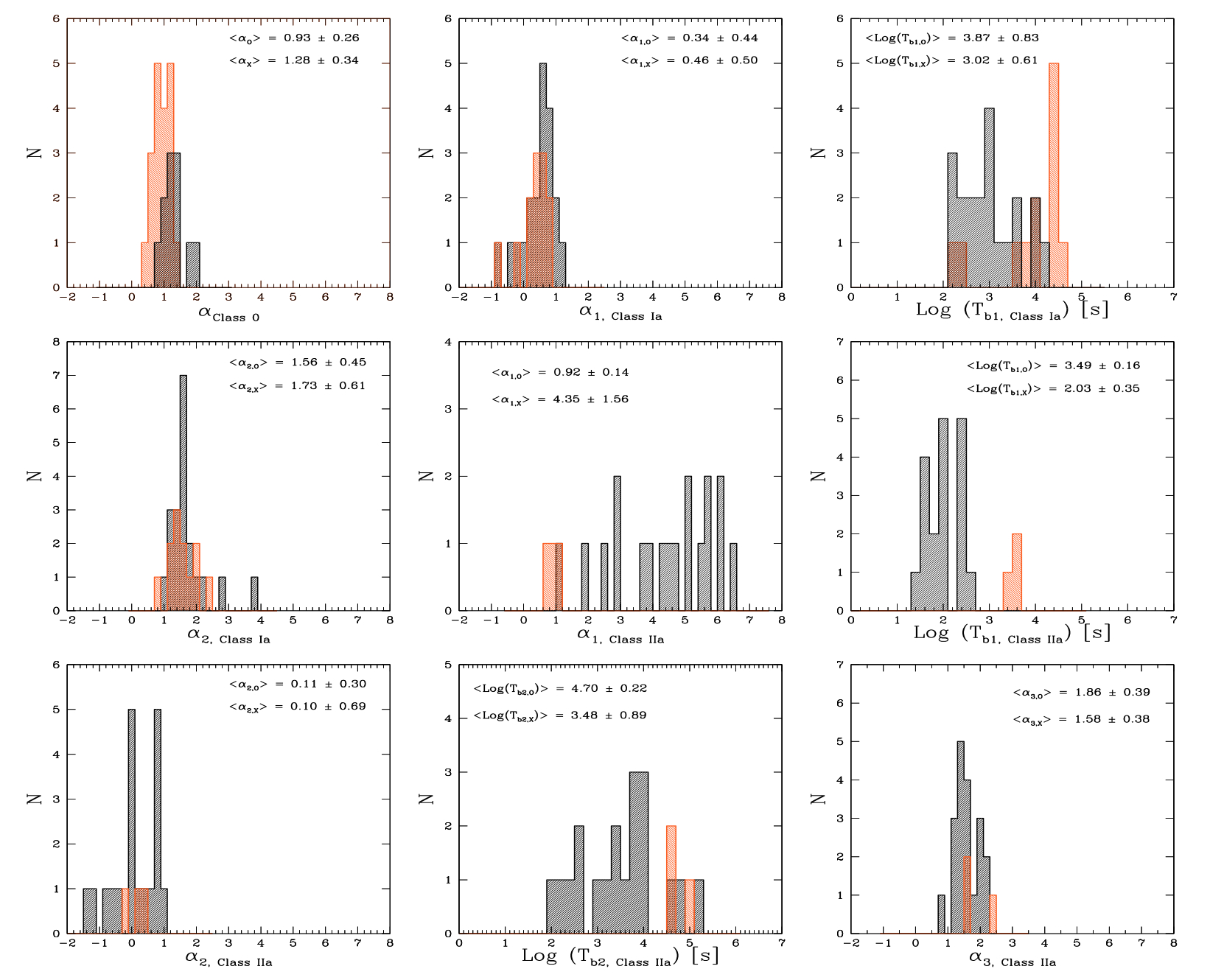}
 \caption{Statistics of the decay indices and break times of the X-ray (black histograms) and optical (orange histograms) light curves for the classes 0, Ia and IIa present in the complete BAT6 sample. We do not display the classes Ib and IIb since the count less than 5 events each. In each single panel we report also the average and the standard deviation for the plotted parameters; break times ($t_{\rm b1}$ and $t_{\rm b2}$) are in order of 10$^{3}$ seconds (as reported in Table \ref{tablc} and \ref{tablc2}).}
\label{Figstat}
\end{figure*}

The comparison of the rest-frame decay indices and the break times for the different classes in the X-ray and optical bands is shown in Fig. \ref{Figstat}. We can observe few peculiarities:

\begin{itemize}
\item For the class 0 (single power-law) the {average} decay index of the X-ray light curves is steeper than the decay of the optical one. However, the two distributions are still consistent within the errors; 
\item The distributions of the decay indices of the class Ia (single broken power-law) are in good agreement for the two bands (both $\alpha_{\rm 1,X, Ia} \simeq \alpha_{\rm 1,O, Ia}$ and $\alpha_{\rm 2,X, Ia} \simeq \alpha_{\rm 2,O, Ia}$), however the change of slope happens at later times in the optical band with respect to the X-ray band ($t_{\rm b1,O, Ia}$ $\gg$ $t_{\rm b1,X, Ia}$);
\item For the class IIa (double broken power-law) the initial decay is always steeper in the X-ray ($\alpha_{\rm 1,X, IIa} \gg \alpha_{\rm 1,O, IIa}$) and the change of slope from steep-to-shallow happens earlier in that band, $\sim$ 1 order of magnitude in time ($t_{\rm b1,X, IIa}$ $\ll$ $t_{\rm b1,O, IIa}$);
\item The flat phase of the class IIa (plateau) is similar for the two bands ($\alpha_{\rm 2,X, IIa}  \simeq \alpha_{\rm 2,O, IIa}$);
\item The late decay indices and break times for the class IIa are in agreement within the errors ($\alpha_{\rm 3,X, IIa} \simeq \alpha_{\rm 3,O, IIa}$) while, as for the early time break time, the shallow-to-steep change of the slope in the optical band happens at later times ($t_{\rm b2,X, IIa}$ $\ll$ $t_{\rm b2,O, IIa}$).
\end{itemize}

For completeness, even if we do not show the histograms for the class Ib in Fig. \ref{Figstat}, the decay indices (pre- and post-break) in the X-ray band for class Ib are steeper than the ones observed in the optical band ($\alpha_{\rm 1,X, Ib} \gg \alpha_{\rm 1,O, Ib}$ and $\alpha_{\rm 2,X, Ib} \gg \alpha_{\rm 2,O, Ib}$), while the break time in the two bands are similar ($t_{\rm b1,O, Ib}$ $\simeq$ $t_{\rm b1,X, Ib}$). The case of class IIb is seen only for one event in the X-rays (GRB~090926B), while the class III defined in Margutti et al. (2013) is never seen in our complete sample. The optical light curve display a clear peak at early times for $\sim 20\%$ of the event in the BAT6 sample.

\subsection{X-ray-optical comparison}

The comparison between the light curves revealed that for only 27$\%$ of the cases the classification is the same in the two bands (larger than what found by \cite{elena}). The power-law decays for the majority of those events are consistent between the two bands and the decay indices are even more in agreement when fewer breaks in the light curves are observed. This is an indication of a possible common (and unique) origin for the observed emission at these wavelengths for these events. For the remaining (large) fraction of GRBs in the BAT6 sample the behaviour is more complex and the agreement between the two bands is more difficult. Those are the cases where additional components (e.g., tail of the prompt emission, flares, energy injection) are seen superposed with the late times afterglow emission. This is a strong indication of different emitting regions that contribute in shaping the observed light curves. Similar contribution of these components for long periods is the cause of the complex behaviours observed when comparing X-ray and optical light curves (\cite{ghise}).

 \begin{table*}
       \caption[]{X-ray light curves fit results: values of $\alpha$ and $t$ represent the decay indices and the break times of the light curves, respectively.  N$_{\rm PL}$ and N$_{\rm BPL}$ are the normalisations of the fitting functions, $\chi^{2}_{red}$ is the goodness of the fit and in the last column we report the classification type of each single GRB following the scheme described in section 2. \\ ${\it a}$ - we did not fit the early time steep exponential decay due to high latitude emission.}
         \label{tablc}
\scalebox{0.78}{
  \begin{tabular}{cccccccccc}
  GRB & $\alpha_{\rm 1}$ & $t_{\rm b1}$ & $\alpha_{\rm 2}$ & $t_{\rm b2}$ & $\alpha_{\rm 3}$ & N$_{\rm PL}$ & N$_{\rm BPL}$ & $\chi^{2}_{red}$ (d.o.f.) & Type \\
   \hline
   & -- & [$\times 10^{3}$  s] & -- & [$\times 10^{3}$ s] & -- & [erg s$^{-1}$] & [erg s$^{-1}$]& -- &\\
  \hline
  050318 & 1.07 $\pm$ 0.25 & 9.7$\pm$ 6.2 & 2.74 $\pm$ 0.58 & -- & -- & -- & 0.00029 $\pm$ 0.00035 & 1.16 (79) & Ia \\
  050401 & 0.35 $\pm$ 0.06 & 1.2 $\pm$ 0.2 & 1.68 $\pm$ 0.09 & -- & -- & -- & 0.053 $\pm$ 0.010 & 1.14 (262) & Ia \\
  050416A & 0.36 $\pm$ 0.18 & 1.2 $\pm$ 1.5 & 1.02 $\pm$ 0.08 & -- & -- & -- & 0.00025 $\pm$ 0.00023 & 1.06 (92) & Ia \\
  050525A & 0.79 $\pm$ 0.11 & 1.2 $\pm$ 0.7 & 1.63 $\pm$ 0.06 & -- & -- & -- & 0.0017 $\pm$ 0.0013 & 1.16 (33) & Ia \\
  050802 & 1.16 $\pm$ 0.07 & 0.3 $\pm$ 0.1 & 0.09 $\pm$ 0.28 & 2.3 $\pm$ 0.5 & 2.19 $\pm$ 0.15 & 5.02 $\pm$ 3.69 & 0.005 $\pm$ 0.001 & 0.91 (153) & IIa\\ 
  050922C & 0.90 $\pm$ 0.07 & 1.9 $\pm$ 1.4 & 1.69 $\pm$ 0.12 & -- & -- & -- & 0.0046 $\pm$ 0.0044 & 1.09 (146) & Ia \\
  060206 & 0.90 $\pm$ 0.04 & 0.2 $\pm$  0.1 & -3.86 $\pm$ 0.93 & 0.7 $\pm$ 0.1 & 1.44 $\pm$ 0.08 & 2.00 $\pm$ 0.43 & 0.077 $\pm$ 0.009 & 1.52 (121) & 0+B\\
  060210 & 2.90 $\pm$ 0.56 & 0.03 $\pm$ 0.01 & 0.72 $\pm$ 0.09 & 9.0 $\pm$ 5.3 & 1.65 $\pm$ 0.12 & (1.2 $\pm$ 0.2)$\times 10^{4}$ & 0.0084 $\pm$ 0.0058 & 1.48 (567) & IIa+FF\\
  060306 & 3.87 $\pm$ 0.50 & 0.05 $\pm$ 0.01 & 0.22 $\pm$ 0.17 & 1.1 $\pm$ 0.7 & 1.31 $\pm$ 0.10 & (1.1 $\pm$ 0.2)$\times 10^{6}$ & 0.023 $\pm$ 0.012 & 1.39 (96) & IIa\\
  060614$^{a}$ & -- & -- & -0.04 $\pm$ 0.09 & 42.5 $\pm$ 5.9 & 1.94 $\pm$ 0.08 & -- & (1.77 $\pm$ 0.25)$\times 10^{-6}$ & 1.80 (154) & IIa\\
  060814 & 2.90 $\pm$ 0.06 & 0.45 $\pm$ 0.20 & -0.43 $\pm$ 0.21& 2.0 $\pm$ 0.3 & 1.38 $\pm$ 0.04 & (1.4 $\pm$ 0.4)$\times 10^{5}$ & 0.0071 $\pm$ 0.0008 & 1.33 (338) & IIa\\
  060908 & -0.02 $\pm$ 0.21& 0.17 $\pm$ 0.06 & 1.63 $\pm$ 0.09 & -- & -- & -- & 0.051 $\pm$ 0.017 & 1.46 (31) & Ia\\
  060912A & 1.07 $\pm$ 0.02 & -- & -- & -- & -- & 0.29 $\pm$ 0.04 & -- & 1.13 (35) & 0\\
  060927 & 0.57 $\pm$ 0.11 & 1.0 $\pm$ 0.3 & 3.79 $\pm$ 1.43 & -- & -- & -- & 0.063 $\pm$ 0.035 & 0.90 (14) & Ia\\
  061007 & 1.73 $\pm$ 0.01 & -- & -- & -- & -- & 1004.0 $\pm$ 41.7 & -- & 1.25 (764) & 0\\
  061021 & 2.60 $\pm$ 0.35 & 0.22 $\pm$ 0.10 & 0.45 $\pm$ 0.16 & 6.2 $\pm$ 5.8 & 1.16 $\pm$ 0.07 & 129.6 $\pm$ 183.2 & (6.2 $\pm$ 4.8)$\times 10^{-5}$ & 1.63 (333) & IIa\\
  061121 & 6.19 $\pm$ 0.09 & 0.10 $\pm$ 0.05 & 0.06 $\pm$ 0.05 & 1.7 $\pm$ 0.2 & 1.48 $\pm$ 0.03 & (1.1 $\pm$ 0.4)$\times 10^{10}$ & 0.013 $\pm$ 0.001 & 1.95 (360) &IIa\\
  061222A & 6.43 $\pm$ 1.43 & 0.09 $\pm$ 0.04 & 0.58 $\pm$ 0.03 & 10.6 $\pm$ 0.2 & 1.48 $\pm$ 0.03 & (1.8 $\pm$ 0.9)$\times 10^{10}$ & 0.0063 $\pm$ 0.0013 & 1.34 (431) & IIa\\
  070306 & 5.68 $\pm$ 0.10 & 0.23 $\pm$ 0.05 & -0.07 $\pm$ 0.06 & 11.9 $\pm$ 1.1 & 2.02 $\pm$ 0.07 & (1.8 $\pm$ 0.8)$\times 10^{10}$ & 0.0023 $\pm$ 0.0002 & 1.27 (240) & IIa\\
  070521 & 0.52 $\pm$ 0.04 & 4.7 $\pm$ 0.8 & 2.20 $\pm$ 0.13 & -- & -- & -- & 0.0022 $\pm$ 0.005 & 1.26 (82) & Ia\\
  071020 & 1.11 $\pm$ 0.01 & -- & -- & -- & -- & 9.98 $\pm$ 0.54 & -- & 2.24 (190) & 0\\
  071112C & 1.33 $\pm$ 0.02 & 0.15 $\pm$ 0.03 & -2.99 $\pm$ 1.12 & 0.33 $\pm$ 0.05 & 3.02 $\pm$ 0.73 & 3.82 $\pm$ 0.32 & 0.0048 $\pm$ 0.0006 & 1.31 (117) & 0+B\\
  071117 & 0.88 $\pm$ 0.06 & -- & -- & -- & -- & 0.26 $\pm$ 0.13 & -- & 1.48 (27) & 0\\
  080319B & 1.15 $\pm$ 0.05 & 0.38 $\pm$ 0.12 & 1.98 $\pm$ 0.04 & -- & -- & -- & 0.26 $\pm$ 0.13 & 1.35 (1624) & Ia+B\\
  080319C & -0.83 $\pm$ 0.37& 0.18 $\pm$ 0.03 & 1.49 $\pm$ 0.03 & -- & -- & -- & 0.68 $\pm$ 0.07 & 1.45 (54) & Ia\\
  080413B & 0.16 $\pm$ 0.12 & 0.24 $\pm$ 0.07 & 1.15 $\pm$ 0.03 & -- & -- & -- & 0.011 $\pm$ 0.002 & 1.25 (217) & Ia\\
  080430 & 1.90 $\pm$  0.12 & 0.3 $\pm$ 0.1 & -0.05 $\pm$ 0.12 & 7.2 $\pm$ 2.1 & 1.13 $\pm$ 0.05 & 8.75 $\pm$ 3.89 & 0.00016 $\pm$ 0.00003 & 1.15 (160) & IIa\\
  080602 & 3.41 $\pm$ 1.22 & 0.08 $\pm$ 0.02 & -3.99 $\pm$ 1.98 & 0.10 $\pm$ 0.09 & 0.67 $\pm$ 0.14 & (2.1 $\pm$ 1.0)$\times 10^{5}$ & 0.026 $\pm$ 0.004 & 0.83 (66) & IIa\\
  080603B & 3.61 $\pm$ 0.21 & 0.06 $\pm$ 0.02 & -1.33 $\pm$ 1.06 & 0.10 $\pm$ 0.06 & 1.12 $\pm$ 0.24 & (4.9 $\pm$ 3.1)$\times 10^{4}$ & 0.10 $\pm$ 0.02 & 1.19 (89) & IIa\\
  080605 & 0.56 $\pm$ 0.05 & 0.48 $\pm$ 0.13 & 1.64 $\pm$ 0.05 & -- & -- & -- & 0.063 $\pm$ 0.018 & 1.10 (312) & Ia\\
  080607 & 5.20 $\pm$ 0.63 & 0.05 $\pm$ 0.02 & 0.78 $\pm$ 0.24 & 0.38 $\pm$ 0.33 & 1.66 $\pm$ 0.09 & (8.2 $\pm$ 1.6)$\times 10^{7}$ & 0.095 $\pm$ 0.106 & 1.51 (334) & IIa+F\\
  080721 & 0.66 $\pm$ 0.01 & 0.72 $\pm$ 0.07 & 1.71 $\pm$ 0.02 & -- & -- & -- & 0.235 $\pm$ 0.026 & 1.12 (1407) & Ia\\
  080804 & 1.11 $\pm$ 0.01 & -- & -- & -- & -- & 5.98 $\pm$ 0.34 & -- & 0.89 (101) & 0\\
  080916A & 2.95 $\pm$ 0.19 & 0.50 $\pm$ 0.20 & 1.21 $\pm$ 0.08 & -- & -- & 3151.0 $\pm$ 2257 & (1.6 $\pm$ 0.3)$\times 10^{-4}$ & 1.36 (135) & Ib+B\\
  081007 & 5.77 $\pm$ 0.79 & 0.12 $\pm$ 0.05 & 0.74 $\pm$ 0.05 & 135.6 $\pm$ 96.2 & 1.68 $\pm$ 0.31& (2.3 $\pm$ 0.8)$\times 10^{8}$ & (4.2 $\pm$ 3.3)$\times 10^{-6}$ & 1.49 (80) & IIa\\
  081121 & 1.47 $\pm$ 0.01 & -- & -- & -- & -- & 956.9 $\pm$ 118.6 & -- & 1.06 (145) & 0\\
  081203A & 5.58 $\pm$ 0.23 & 0.05 $\pm$ 0.01 & 1.02 $\pm$ 0.04 & 4.4 $\pm$ 1.3 & 2.29 $\pm$ 0.15 & (2.3 $\pm$ 1.8)$\times 10^{8}$ & 0.0020 $\pm$ 0.0009 & 1.22 (321) & IIa\\
  081221 & 5.03 $\pm$ 0.66 & 0.09 $\pm$ 0.02 & -0.75 $\pm$ 0.68 & 0.14 $\pm$ 0.03 & 1.42 $\pm$ 0.11 & (3.9 $\pm$ 1.0)$\times 10^{8}$ & 0.34 $\pm$ 0.03 & 1.22 (321) & IIa+B\\
  081222 & 0.87 $\pm$ 0.04 & 3.3 $\pm$ 1.6 & 1.81 $\pm$ 0.34 & -- & -- & -- & 0.0073 $\pm$ 0.0043 & 1.06 (372) & Ia+B\\
  090102 & -0.32 $\pm$ 0.31 & 0.25 $\pm$ 0.03 & 1.50 $\pm$ 0.03 & -- & -- & -- & 0.095 $\pm$ 0.009 & 1.16 (139) & Ia\\
  090201 &  1.19 $\pm$ 0.01 & -- & -- & -- & -- & 74.8 $\pm$ 10.2 & -- & 1.29 (166) & 0\\ 
  090424 & 0.66 $\pm$ 0.04 & 2.0 $\pm$ 0.9 & 1.31 $\pm$ 0.04 & -- & -- & -- & 0.0025 $\pm$ 0.0011 & 1.84 (689) & Ia\\
  090709A & 4.52 $\pm$ 0.21 & 0.07 $\pm$ 0.02 & -1.13 $\pm$ 0.10 & 0.21 $\pm$ 0.02 & 1.49 $\pm$ 0.02 & (6.7 $\pm$ 5.2)$\times 10^{6}$ & 0.43 $\pm$ 0.01 & 1.53 (445) & IIa\\ 
  090715B & 1.31 $\pm$ 0.05 & -- & -- & -- & -- & 81.7 $\pm$ 32.5 & -- & 1.64 (61) & 0+FFB\\
  090812 & 3.39 $\pm$ 0.76 & 0.91 $\pm$ 0.12 & 1.17 $\pm$ 0.02 & -- & -- & (3.0 $\pm$ 0.7)$\times 10^{4}$ & 29.0 $\pm$ 1.1 & 1.28 (426) & Ib+FF\\
  090926B & -1.23 $\pm$ 0.87 & 0.05 $\pm$ 0.01 & 2.79 $\pm$ 0.26 & 2.8 $\pm$ 0.4 & 1.02 $\pm$ 0.09 & 1.55 $\pm$ 1.05 & 0.45 $\pm$ 0.06 & 1.00 (174) & IIb\\
  091018 & -0.16 $\pm$ 0.13 & 0.16 $\pm$ 0.03 & 1.29 $\pm$ 0.02 & -- & -- & -- & 0.022 $\pm$ 0.003 & 1.32 (136) & Ia\\
  091020 & 6.07 $\pm$ 0.49 & 0.05 $\pm$ 0.02 & 0.87 $\pm$ 0.06 & 5.7 $\pm$ 4.0 & 1.61 $\pm$ 0.10 & (2.3 $\pm$ 3.9)$\times 10^{8}$ & 0.0013 $\pm$ 0.0009 & 1.09 (260) & IIa\\
  091127 & 0.80 $\pm$ 0.10 & 16.1 $\pm$ 8.7 & 1.65 $\pm$ 0.06 & -- & -- & -- & 0.00036 $\pm$ 0.00025 & 1.44 (368) & Ia\\
  091208B & 0.16 $\pm$ 0.21 & 0.60 $\pm$ 0.30 & 1.27 $\pm$ 0.05 & -- & -- & -- & 0.0046 $\pm$ 0.0021 & 1.26 (61) & Ia\\
  100615A & 4.24 $\pm$ 0.19 & 0.08 $\pm$  0.02 & -0.64 $\pm$ 0.28 & 0.32 $\pm$ 0.07 & 0.82 $\pm$ 0.03 & (1.5 $\pm$ 1.0)$\times 10^{6}$ & 0.027 $\pm$ 0.002 & 1.02 (107) & IIa\\ 
  100621A & 4.68 $\pm$ 0.15 & 0.3 $\pm$ 0.1 & 0.72 $\pm$ 0.03 & 54.3 $\pm$ 13.3 & 1.92 $\pm$ 0.10 & (2.3 $\pm$ 1.7)$\times 10^{8}$ & (6.0 $\pm$ 1.8)$\times 10^{-5}$ & 1.33 (278) & IIa\\
  100728B &1.98 $\pm$ 0.28 & -- & -- & -- & -- & 80.1 $\pm$ 8.5 & -- & 0.96 (33) & 0+B\\
  110205A & 6.47 $\pm$ 0.25 & 0.15 $\pm$ 0.08 & 1.68 $\pm$ 0.01 & -- & -- & -- & 1040.9 $\pm$ 95.6 & 1.25 (682) & Ib+F\\
  110503A & 0.95 $\pm$ 0.04 & 8.0 $\pm$ 6.3 & 1.53 $\pm$ 0.08 & -- & -- & -- & 0.0012 $\pm$ 0.0008 & 1.05 (394) & Ia\\
  \hline
  \end{tabular}
  }
\end{table*}


 \begin{table*}
       \caption[]{Optical light curves fit results: values of $\alpha$ and $t$ represent the decay indices and the break times of the light curves, respectively.  N$_{\rm PL}$ and N$_{\rm BPL}$ are the normalisations of the fitting functions, $\chi^{2}_{red}$ is the goodness of the fit and in the last column we report the classification type of each single GRB following the scheme described in section 2. \\ {\it a} - we did not fit the late time host galaxy and/or supernovae component.}
         \label{tablc2}
\scalebox{0.73}{
  \begin{tabular}{cccccccccccc}
  GRB & $\alpha_{\rm 1}$ & $t_{\rm b1}$ & $\alpha_{\rm 2}$ & $t_{\rm b2}$ & $\alpha_{\rm 3}$ & $t_{\rm b3}$ & $\alpha_{\rm 4}$ & N$_{\rm PL}$ & N$_{\rm BPL}$ & $\chi^{2}_{red}$ (d.o.f.) & Type \\
   \hline
   & -- & [$\times 10^{3}$  s] & -- & [$\times 10^{3}$ s] & -- & [$\times 10^{3}$ s] & -- & [erg s$^{-1}$] & [erg s$^{-1}$]& -- & \\
  \hline
  050318 & 1.02 $\pm$ 0.11 & -- & -- & -- & -- & -- & -- & 464.5 $\pm$ 41.0 & -- & 0.96 (6) & 0 \\
  050401 & 0.77 $\pm$ 0.02 & -- & -- & -- & -- & -- & -- & 2.52 $\pm$ 0.32 & -- & 1.49 (36) & 0 \\
  050416A & 0.35 $\pm$ 0.27 & 48.5 $\pm$ 48.3 & 0.73 $\pm$ 0.34 & -- & -- & -- & -- & -- & 0.007 $\pm$ 0.037 & 13.6 (38) & Ia \\
  050525A$^{a}$ & 0.83 $\pm$ 0.06 & 8.7 $\pm$ 4.9 & 1.64 $\pm$ 0.09 & -- & -- & -- & -- & -- & 0.32 $\pm$ 0.21 & 1.21 (93) & Ia \\
  050802 & 0.77 $\pm$ 0.02 & -- & -- & -- & -- & -- & -- & 21.8 $\pm$ 3.3 & -- & 2.38 (33) & 0\\ 
  050922C & 0.71 $\pm$ 0.05 & 5.4 $\pm$ 3.1 & 1.61 $\pm$ 0.15 & -- & -- & -- & -- & -- & 0.19 $\pm$  0.12 & 3.17 (57) & Ia \\
  060206 & 1.17 $\pm$ 0.45 & 0.4 $\pm$  0.1 & -21.3 $\pm$ 1.96 & 0.63 $\pm$ 0.01 & 1.14 $\pm$ 0.06 & 5.6 $\pm$ 1.2 & 2.24 $\pm$ 0.33 & 54.9 $\pm$ 8.1 & 0.68 $\pm$ 0.02 & 4.06 (222) & 0+B\\
  060210 & -0.15 $\pm$ 0.24 & 0.14 $\pm$ 0.05 & 1.34 $\pm$ 0.11 & -- & -- & -- & -- & -- & 0.26 $\pm$ 0.07 & 1.00 (33) & Ia\\
  060306 & -- & -- & -- & -- & -- & --& -- & -- & -- & -- & --\\
  060614$^{a}$ & -- & -- & -0.29 $\pm$ 0.08 & 50.4 $\pm$ 4.9 & 1.93 $\pm$ 0.09 & -- & -- & -- & 0.106 $\pm$ 0.009 & 1.68 (35) & Ia\\
  060814 & -- & -- & -- & -- & -- & -- & -- & --& -- & -- & --\\
  060908 & 1.08 $\pm$ 0.01& -- & -- & -- & -- & -- & -- & 176.7 $\pm$ 11.9 & -- & 2.31 (60) & 0\\
  060912A & 1.07 $\pm$ 0.05 & 4.50 $\pm$ 0.30 & -0.20 $\pm$ 0.24 & 88.4 $\pm$ 52.1 & 1.62 $\pm$ 0.46 & -- & -- & 106.7 $\pm$ 22.8 & 0.019 $\pm$ 0.008 & 0.87 (26) & IIa\\
  060927 & 1.18 $\pm$ 0.13 & -- & -- & -- & -- & -- & -- & 13.8 $\pm$ 9.1 & -- & 1.57 (13) & 0\\
  061007 & -3.86 $\pm$ 0.19 & 0.033 $\pm$ 0.007 & 1.67 $\pm$ 0.01 & -- & -- & -- & -- & -- & 562.1 $\pm$ 15.0 & 2.85 (105) & 0+Onset\\
  061021 & 0.90 $\pm$ 0.06 & 3.2 $\pm$ 0.21 & 0.14 $\pm$ 0.12 & 42.3 $\pm$ 8.3 & 2.32 $\pm$ 0.36 & -- & -- & 96.3 $\pm$ 26.6 & 0.043 $\pm$ 0.009 & 1.38 (25) & IIa\\
  061121 & 0.79 $\pm$ 0.91 & 0.04 $\pm$ 0.01 & -7.00 $\pm$ 0.71 & 0.07 $\pm$ 0.01 & 0.78 $\pm$ 0.42  & -- & -- & 13.3 $\pm$ 4.3 & -- & 1.54 (69) & 0+B\\
  061222A & 0.44 $\pm$ 0.11 & -- & -- & -- & -- & -- & -- & -- & -- & -- (0) & 0\\
  070306 & -0.79 $\pm$ 0.27 & 26.7 $\pm$ 6.0 & 2.33 $\pm$ 0.45 & -- & -- & -- & -- & -- & 0.12 $\pm$ 0.03 & 1.19 (1) & Ia\\
  070521 & -- & -- & -- & -- & -- & --& -- & --& -- & -- & --\\
  071020$^{a}$ & 1.08 $\pm$ 0.06 & -- & -- & -- & -- & -- & -- & 111.3 $\pm$ 38.3 & -- & 6.41 (2) & 0\\
  071112C & -3.26 $\pm$ 2.08 & 0.07 $\pm$ 0.01 & 0.87 $\pm$ 0.02 & -- & -- & -- & -- & -- & 0.97 $\pm$ 0.10 & 1.52 (23) & 0+Onset\\
  071117 & 0.83 $\pm$ 0.59 & -- & -- & -- & -- & -- & -- & 4.13 $\pm$ 2.38 & -- & 3.44 (4) & 0\\
  080319B & -4.26 $\pm$ 0.54 & 0.009 $\pm$ 0.002 & 2.25 $\pm$ 0.02 & 0.58 $\pm$ 0.03 & 1.18 $\pm$ 0.01 & -- & -- & (6.6 $\pm$ 0.3)$\times 10^{4}$ & 3.23 $\pm$ 0.30 & 2.55 (175) & Ib+Onset\\
  080319C & 0.68 $\pm$ 0.03& 0.06 $\pm$ 0.02 & -6.45 $\pm$ 1.31 & 0.09 $\pm$ 0.02 & 1.40 $\pm$ 0.09 & -- & -- & 3.19 $\pm$ 0.44 & -- & 2.53 (15) & 0+B\\
  080413B & 0.79 $\pm$ 0.05 & 2.10 $\pm$ 0.30 & -0.41 $\pm$ 0.11 & 34.2 $\pm$ 5.3 & 1.64 $\pm$ 0.11 & -- & -- & 26.2 $\pm$ 7.1 & 0.118 $\pm$ 0.011 & 2.66 (51) & IIa \\
  080430 & 0.49 $\pm$  0.06 & 22.9 $\pm$ 2.3 & 1.37 $\pm$ 0.41 & -- & -- & -- & -- & -- & 0.046 $\pm$ 0.039 & 1.11 (18) & Ia\\
  080602 & -- & -- & -- & -- & -- & --& -- & --& -- & -- & --\\
  080603B & 0.60 $\pm$ 0.02 & 9.2 $\pm$ 1.5 & 2.07 $\pm$ 0.14 & -- & -- & -- & -- & -- & 0.108 $\pm$ 0.012 & 1.35 (17) & Ia\\
  080605 & 0.60 $\pm$ 0.01 & -- & -- & -- & -- & -- & -- & 4.86 $\pm$ 3.78 & -- & 1.47 (35) & 0\\
  080607 & 1.45 $\pm$ 0.03 & 0.45 $\pm$ 0.05 & -1.73 $\pm$ 0.88 & 0.72 $\pm$ 0.17 & 2.32 $\pm$ 0.58 & -- & -- & 110.0 $\pm$ 14.8 & 0.041 $\pm$ 0.006 & 3.18 (40) & 0+B\\
  080721 & 1.18 $\pm$ 0.01 & -- & -- & -- & -- & -- & -- & 1053.7 $\pm$ 85.0 & -- & 1.27 (30) & 0\\
  080804 & -2.53 $\pm$ 0.68 & 0.020 $\pm$ 0.003 & 0.86 $\pm$ 0.01 & -- & -- & -- & -- & -- & 2.13 $\pm$ 0.18 & 1.12 (12) & 0+Onset\\
  080916A & 0.58 $\pm$ 0.09 & -- & -- & -- & -- & -- & -- & 35.6 $\pm$ 18.3 & -- & 3.38 (12) & 0\\
  081007 & -4.12 $\pm$ 1.92 & 0.08 $\pm$ 0.01 & 2.67 $\pm$ 0.39 & 0.16 $\pm$ 0.03 & 0.65 $\pm$ 0.16 & -- & -- & 0.49 $\pm$ 0.08 & 3.28 $\pm$ 0.19 & 1.78 (47) & Ib+Onset\\
  081121 & 1.19 $\pm$ 0.04 & -- & -- & -- & -- & -- & -- & (2.4 $\pm$ 0.5)$\times 10^{3}$ & -- & 1.64 (1) & 0\\
  081203A & -2.19 $\pm$ 0.07 & 0.10 $\pm$ 0.01 & 1.55 $\pm$ 0.01 & -- & -- & -- & -- & -- & 77.4 $\pm$ 1.4 & 1.05 (54) & 0+Onset\\
  081221 & -- & -- & -- & -- & -- & --& -- & --& -- & -- & --\\
  081222 & 1.18 $\pm$ 0.02 & -- & -- & -- & -- & -- & -- & (1.1 $\pm$ 0.1)$\times 10^{3}$ & -- & 1.14 (20) & 0\\
  090102 & 2.09 $\pm$ 0.21 & 0.43 $\pm$ 0.08 & 1.00 $\pm$ 0.01 & -- & -- & -- & -- & -- & 65.0 $\pm$ 8.4 & 1.69 (64) & Ib\\
  090201 & -- & -- & -- & -- & -- & --& -- & --& -- & -- & --\\
  090424$^{a}$ & 2.35 $\pm$ 0.40 & 0.08 $\pm$ 0.05 & 0.86 $\pm$ 0.03 & -- & -- & -- & -- & -- & 200.2 $\pm$ 50.7 & 1.79 (17) & Ib\\
  090709A & 0.96 $\pm$ 0.02 & -- & -- & -- & -- & -- & -- & 33.9 $\pm$ 3.9 & -- & 0.18 (4) & 0\\ 
  090715B & 0.22 $\pm$ 0.02 & 29.3 $\pm$ 5.0 & 1.40 $\pm$ 0.09 & -- & -- & -- & -- & -- & 0.016 $\pm$ 0.001 & 2.23 (40) & Ia\\
  090812 & -2.85 $\pm$ 0.93 & 0.014 $\pm$ 0.001 & 1.26 $\pm$ 0.03 & -- & -- & -- & -- & -- & 6.85 $\pm$ 0.56 & 3.76 (15) & 0+Onset\\
  090926B & -- & -- & -- & -- & -- & --& -- & --& -- & -- & --\\
  091018 & 0.69 $\pm$ 0.09 & 3.3 $\pm$ 7.0 & 1.14 $\pm$ 0.10 & -- & -- & -- & -- & -- & 0.38 $\pm$ 0.85 & 1.05 (68) & Ia\\
  091020 & 0.24 $\pm$ 0.40 & 0.22 $\pm$ 0.26 & 1.24 $\pm$ 0.12 & -- & -- & -- & -- & -- & 2.20 $\pm$ 2.42 & 4.62 (35) & Ia\\
  091127$^{a}$ & 0.33 $\pm$ 0.01 & 24.3 $\pm$ 1.6 & 1.76 $\pm$ 0.06 & -- & -- & -- & -- & -- & 0.51 $\pm$ 0.03 & 1.13 (187) & Ia\\
  091208B & 0.60 $\pm$ 0.02 & 26.1 $\pm$ 3.3 & 2.08 $\pm$ 0.44 & -- & -- & -- & -- & -- & 0.017 $\pm$ 0.003 & 1.84 (25) & Ia\\
  100615A & -- & -- & -- & -- & -- & --& -- & --& -- & -- & --\\
  100621A & -2.48 $\pm$ 2.04 & 0.29 $\pm$  0.03 & 1.04 $\pm$ 0.14 & 2.3 $\pm$ 0.2 & -20.0 $\pm$ 2.5 & 2.99 $\pm$ 0.03 & 1.22 $\pm$ 0.07 & 1.03 $\pm$ 0.06 & 1.25 $\pm$ 0.18 & 7.73 (44) & 0+Onset+B\\
  100728B &-4.83 $\pm$ 6.11 & 0.011 $\pm$ 0.001 & 0.93 $\pm$ 0.02 & -- & -- & -- & -- & -- & 3.67 $\pm$ 0.35 & 2.55 (23) & 0+Onset\\
  110205A & -7.29 $\pm$ 0.68 & 0.25 $\pm$ 0.03 & 1.47 $\pm$ 0.01 & -- & -- & -- & -- & -- & 8.30 $\pm$ 0.17 & 1.53 (102) & 0+Onset\\
 110503A & 0.83 $\pm$ 0.02 & -- & -- & -- & -- & -- & -- & 71.3 $\pm$ 8.9 & -- & 1.09 (21) & 0\\
  \hline
  \end{tabular}
}
\end{table*}

\section{Conclusions}

We investigated the rest-frame optical properties of the complete BAT6 sample. We unarguably demonstrate that the optical luminosity at $t_{\rm rf}$ = 12 hr has a uniform distribution around a mean value Log(L$_{\rm R}$) = 29.9 (dispersion $\sigma$ = 0.8). No bimodality is observed, as found in published studies based on incomplete samples. Previous claims of bimodality were based on inhomogenity of the analysis for the determination of the main parameters that influence the estimate of the rest-frame luminosity ($\beta_{\rm O}$ and A$_{\rm V}$). In this work we used spectral and absorption parameters that has been estimated following a consistent procedure (described in details in \cite{covino}) and therefore our result is more robust.

The comparison between optical and X-ray rest-frame light curves revealed that, first of all, the complexity and different behaviours observed in these bands are strongly related to the emission components that might still be relevant in both bands. Only for few cases the observed light curves show similar decay slopes in the same time interval: this is an indication for a common origin of the emission. Instead, for the majority of the events of the sample {\bf ($\sim 70\%$)} this is not the case and the X-ray emission seems to have still a strong contribution from the prompt emission or from some late time central engine activity, whose contribution is negligible (or not detected) in the optical band where the afterglow emission dominates. Second, the distribution of the rest-frame X-ray luminosity is slightly broader (and naturally brighter) with respect to the distribution of the rest-frame optical luminosity while this trend is the opposite when considering the E$_{\rm iso}$-normalised quantities. This again indicates that the emission in the X-ray is a better proxy of the prompt emission (since there is a long-lasting contribution of prompt emission in the X-ray) while the emission in the optical is strongly related to the afterglow emission only.

\begin{acknowledgements}

We thank the anonymous referee for the valuable comments, which significantly contributed to improve the quality of the publication. This research has been supported by ASI grant INAF I/004/11/1. This work made use of data supplied by the UK Swift Science Data Centre at the University of Leicester.

\end{acknowledgements}


\begin{thebibliography}{}

\bibitem[Campana et al. 2012]{campana} Campana, S., Salvaterra, R., Melandri, A. et al., 2012, MNRAS, 421, 1697
\bibitem[Covino et al. 2013]{covino} Covino, S., Melandri A., Salvaterra, R., et al., 2013, MNRAS, 432, 1231 
\bibitem[D'Avanzo et al. 2012]{davanzo} D'Avanzo, P., Salvaterra, R., Sbarufatti, B., et al. 2012, MNRAS, 425, 506
\bibitem[Evans et al. 2009]{evans0} Evans, P., Beardmore, A. P., Page, K. L., et al., 2009, MNRAS, 397, 1177
\bibitem[Evans et al. 2010]{evans} Evans, P., Willingale, R., Osborne, J. P., et al., 2010, A$\&$A, 519, 102
\bibitem[Gehrels et al. 2004]{geh} Gehrels, N., Chincarini, G., Giommi, P., et al. 2004, ApJ, 611, 1005
\bibitem[Gehrels et al. 2008]{geh2} Gehrels, N., Barthelmy, S. D., Burrows, D. N., et al. 2008, ApJ, 689, 1161
\bibitem[Ghirlanda et al. 2012]{ghirla} Ghirlanda, G., Ghisellini, G., Nava, L., et al., 2012, MNRAS, 422, 2553
\bibitem[Ghirlanda et al. 2013]{ghirla2} Ghirlanda, G., Salvaterra, R., Burlon, D., et al. 2013, MNRAS, 435, 2543
\bibitem[Ghisellini et al. 2009]{ghise} Ghisellini, G., Nardini, M., Ghirlanda, G., $\&$ Celotti, A., 2009, MNRAS, 393, 253
\bibitem[Kann et al. 2006]{kann} Kann, D. A., Klose, S., $\&$ Zeh, A. 2006, ApJ, 641, 993
\bibitem[Kann et al. 2011]{kann2} Kann, D. A., Klose, S., Zhang, B., et al. 2011, ApJ, 734, 96
\bibitem[Li et al. 2012]{li} Li, L., Liang, E.-W., Tang, Q.-W., et al. 2012, ApJ, 758, 27
\bibitem[Liang $\&$ Zhang 2006]{liang} Liang, E., $\&$ Zhang, B. 2006, ApJ, 638, 67
\bibitem[Liang et al. 2012]{liang2} Liang, E.-W., Li, L., Gao, H., et al. 2012, ApJ, 774, 13 
\bibitem[Mangano et al. 2006]{mangano} Mangano, V., Holland, S. T., Malesani, D., et al. 2006, A$\&$A, 470, 105
\bibitem[Margutti et al. 2013]{raffa} Margutti, R., Zaninoni, E., Bernardini, M. G., et al. 2013, MNRAS, 428, 729
\bibitem[Melandri et al. 2008]{mela0} Melandri, A., Mundell, C. G., Kobayashi, S., et al., 2008, ApJ, 686, 1209
\bibitem[Melandri et al. 2012]{mela} Melandri, A., Sbarufatti, B., D'Avanzo, P., et al., 2012, MNRAS, 421, 1265
\bibitem[Nardini et al. 2006]{nardini1} Nardini, M., Ghisellini, G., Ghirlanda, G., et al. 2006, A$\&$A, 451, 821
\bibitem[Nardini et al. 2008]{nardini2} Nardini, M., Ghisellini, G., $\&$ Ghirlanda, G., 2008, MNRAS, 386, 87
\bibitem[Nava et al. 2012]{nava} Nava, L., Salvaterra, R., Ghirlanda, G., et al., 2012, MNRAS, 421, 1256
\bibitem[Nousek et al. 2006]{nousek} Nousek, J. A., Kouveliotou, C., Grupe, D., et al. 2006, ApJ, 642, 389
\bibitem[Nysewander et al. 2009]{nyse} Nysewander, M., Fruchter, A. S., $\&$ PeÕer, A., 2009, ApJ, 701, 824
\bibitem[Oates et al. 2009]{oates} Oates, S. R., Page, M. J., Schady, P., et al. 2009, MNRAS, 395, 490 
\bibitem[Roming et al. 2009]{roming} Roming, P. W. A., Koch, T. S., Oates, S. R., et al. 2009, ApJ, 690, 163
\bibitem[Rykoff et al. 2009]{rykoff} Rykoff, E. S., Aharonian, F., Akerlof, C. W., et al. 2009, ApJ, 702, 489
\bibitem[Salvaterra et al. 2012]{ruben} Salvaterra, R., Campana, S., Vergani, S. D., et al., 2012, ApJ, 749, 68
\bibitem[Zaninoni et al. 2013]{elena} Zaninoni, E., Bernardini, M. G., Margutti, R., Oates, S., $\&$ Chincarini, G. 2013, A$\&$A, 557, 12
\bibitem[Zhang et al. 2006]{zhang} Zhang, B., Fan, Y. Z., Dyks, J., et al. 2006, ApJ, 642, 354

\end{thebibliography}
\end{document}